\documentclass [superscriptaddress,reprint,amsmath,amssymb,aps,prl,showkeys,floatfix]{revtex4-2}
\usepackage{xcolor}
\usepackage{soul}
\usepackage[normalem]{ulem}
\usepackage[hidelinks]{hyperref}
\usepackage{graphicx}
\usepackage{dcolumn}
\usepackage{bm}
\usepackage{color}
\usepackage{float}
\usepackage[utf8]{inputenc}
\usepackage[mathlines]{lineno}
\usepackage{multirow}
\usepackage{amsmath}
\usepackage{pdfpages}
\usepackage{pgffor}
\makeatletter
\AtBeginDocument{\let\LS@rot\@undefined}
\makeatother
\usepackage{hyperref}
\usepackage{soul,xcolor}
\usepackage{natbib}
\usepackage{amsmath} 
\usepackage{xcolor}
\usepackage{xr}
\usepackage{chemformula}
\usepackage{siunitx}
\usepackage{physics}
\soulregister\cite7
\soulregister\ref7
\setstcolor{blue}
\newcommand{\doHMN}[2]{%
  \begingroup\lccode`~=`#1
  \lowercase{\endgroup\let~}#2%
  \mathcode`#1="8000
}

\makeatletter
\newcommand*{\addFileDependency}[1]{
  \typeout{(#1)}
  \@addtofilelist{#1}
  \IfFileExists{#1}{}{\typeout{No file #1.}}
}
\makeatother

\usepackage{tikz}

\begin{document}
\preprint{APS/123-QED}
\title{Dynamical control of topology in ferroelectric skyrmions \\ via twisted light}
\author{Lingyuan Gao}
\author{Sergei Prokhorenko}
\author{Yousra Nahas}
\author{Laurent Bellaiche}
\email{laurent@uark.edu}
\affiliation{Physics Department and Institute for Nanoscience and Engineering, University of Arkansas, Fayetteville, Arkansas, 72701, USA}

\date{\today}

\begin{abstract}
Twisted light carries a non-zero orbital angular momentum, that can be transferred from light to electrons and particles ranging from nanometers to micrometers. Up to now, the interplay between twisted light with dipolar systems has scarcely been explored, though the latter bear abundant forms of topologies such as skyrmions and embrace strong light-matter coupling. Here, using first-principles-based simulations, we show that twisted light can excite and drive dynamical polar skyrmions and transfer its nonzero winding number to ferroelectric ultrathin films. The skyrmion is successively created and annihilated alternately at the two interfaces, and experiences a periodic transition from a markedly ``Bloch" to ``N\'eel” character, accompanied with the emergence of a “Bloch point” topological defect with vanishing polarization. The dynamical evolution of
skyrmions is connected to a constant jump of topological number between “0” and “1” over time. These intriguing phenomena are found to have an electrostatic origin. Our study thus demonstrates that, and explains why, this unique light-matter interaction can be very powerful in creating and manipulating topological solitons in functional materials.

\end{abstract}

\maketitle

With their noncolinear spin patterns and particle-like features~\cite{nagaosa2013topological,fert2017magnetic}, skyrmions have attracted enormous interests in condensed matter physics. Their unique  properties, such as topological hall effect~\cite{neubauer2009topological,li2013robust} and low current-driven motion~\cite{jiang2015blowing,woo2016observation}, enable promising applications, including racetrack memory and logic gates~\cite{parkin2008magnetic,fert2013skyrmions,zhang2015magnetic,parkin2015memory}. Inspired by these discoveries, scientists recently aimed to seek an electric counterpart of topological solitons in ferroelectric systems, as they can be more easily controlled by electric field~\cite{wang2018ferroelectrically,behera2022electric,zhu2022dynamics}. Polar patterns with skyrmion topologies have been predicted in BaTiO$_3$/SrTiO$_3$ nanocomposites and bulk PbTiO$_3$ by first-principles-based approaches~\cite{nahas2015discovery,gonccalves2018theoretical}, and lately skyrmion-like polar solitons have been observed in (PbTiO$_3$)$_n$/(SrTiO$_3$)$_n$ superlattices and SrTiO$_3$/Pb(Zr$_{x}$Ti$_{1-x}$)O$_3$/SrTiO$_3$ heterostructures~\cite{das2019observation,nahas2020topology,han2022high}.
Different from magnetic skyrmions forming out of asymmetric exchange interaction between spins~\cite{dzyaloshinsky1958thermodynamic,moriya1960anisotropic}, nontrivial ferroelectric structures typically originate from a competition between elastic, electrostatic and gradient energies~\cite{nahas2015discovery,yadav2016observation}.

Moreover, the field of twisted light has rapidly developed over the last thirty years in multiple respects~\cite{allen1992orbital,shen2019optical, rosen2022interplay}. One type of twisted light, optical vortex (OV) beams, With features akin to superfluid vortices, carry a non-zero orbital angular momentum (OAM) with a phase singularity circulated by helical wavefront. Previous works show that the nontrivial field patterns can be imprinted in the generated photocurrent or motions of particles~\cite{he1995direct,paterson2001controlled,macdonald2002creation,grier2003revolution,sederberg2020vectorized,ji2020photocurrent}, and nonuniform heating and magnetic field in the form of a vortex beam have been predicted to induce topological defects in chiral magnets~\cite{fujita2017ultrafast,fujita2017encoding}. Thus it is timely and legitimate to wonder if the topological phase singularity in an optical field carrying OAM can be printed on dipoles to create topological solitons in ferroelectric systems.

Here, we indeed demonstrate that the OV beam can induce dynamical polar skyrmions at interfacial layers in ferroelectric Pb(Zr$_{x}$Ti$_{1-x}$)O$_3$ (PZT) ultrathin films, and allow such topological defects to move in a controllable manner. By interaction between this nontrivial light and electric dipoles, the in-plane electric field can induce a dynamical polar skyrmion with an unusual out-of-plane-dipole-component-switching. A topological defect ``Bloch point" with vanishing polarization is identified for the first time in ferroelectrics and is involved in skyrmion creation~\cite{feldtkeller1965mikromagnetisch, thiaville2003micromagnetic}. Moreover, the robustness of the mechanism manifests itself in the sense that the skyrmion is well maintained under different conditions, and its intrinsic characteristics such as size and switching speed are highly tunable by controlling external variables of the beam.

Technically, we use a first-principles-based effective Hamiltonian approach to study ferroelectric ultrathin films made of Pb(Zr$_{0.4}$Ti$_{0.6}$)O$_3$ (PZT)~\cite{kornev2004ultrathin,ponomareva2005atomistic,ponomareva2005low}. Under different mechanical and electrical conditions, PZT and related systems have  been found to exhibit several exotic phases, including vortex~\cite{gruverman2008vortex}, flux-closure~\cite{tang2015observation}, and nanobubble domains~\cite{zhang2017nanoscale}. Here, we introduce the lowest order of the ``Laguerre-Gaussian" beam (i.e., $l$ = 0)  propagating along the direction normal to the film and passing through the center of each (001) layer to interact with the well-equilibrated monodomain at 10 K (\textbf{Fig. 1})~\cite{allen1992orbital}. As such, the time-dependent, in-plane electric field can be written as:
\begin{align}
    \Vec{E}(\Vec{r},t) = E_0(\frac{\sqrt{2}\rho}{w})^{|m|}e^{-\frac{{\rho}^2}{w^2}}\nonumber \\\big{(}\cos{( m\phi - \omega t)} \Vec{e}_x - \sigma \sin{( m\phi - \omega t)} \Vec{e}_y \big{)}.
\end{align}
Here, $\Vec{e}_x$ and $\Vec{e}_y$ are polarization vectors along the $x$- and $y$-axes that  are lying along the [100] and [010] pseudo-cubic (p.c.) directions, respectively; $m\phi$ and $\sigma$ characterize the phase twist of the field and the handness of the  polarization, corresponding to orbital and spin angular momentum, respectively; $E_0$, $\omega$ and $w$ denote the field magnitude, light frequency and beam radius, respectively. $\Vec{r}$ ($\rho$, $\phi$, $z$) is the position vector from the center of the supercell to any B-site of the film in cylindrical coordinates. In the present work, we set $\omega = 1$ THz, $w$ = 5 unit cells (u.c.), and we consider $\sigma = 1, m = -1$ for simplicity; as such, this electric field $\Vec{E}$ always has an in-plane orientation and carries a winding number $w_v[\Vec{E}] = -m$~\cite{shintani2016spin}. Computational details about other parameters and justifications for choosing their values are detailed in the Supplementary Material.

\begin{figure} [H]
\includegraphics[width = 80mm]{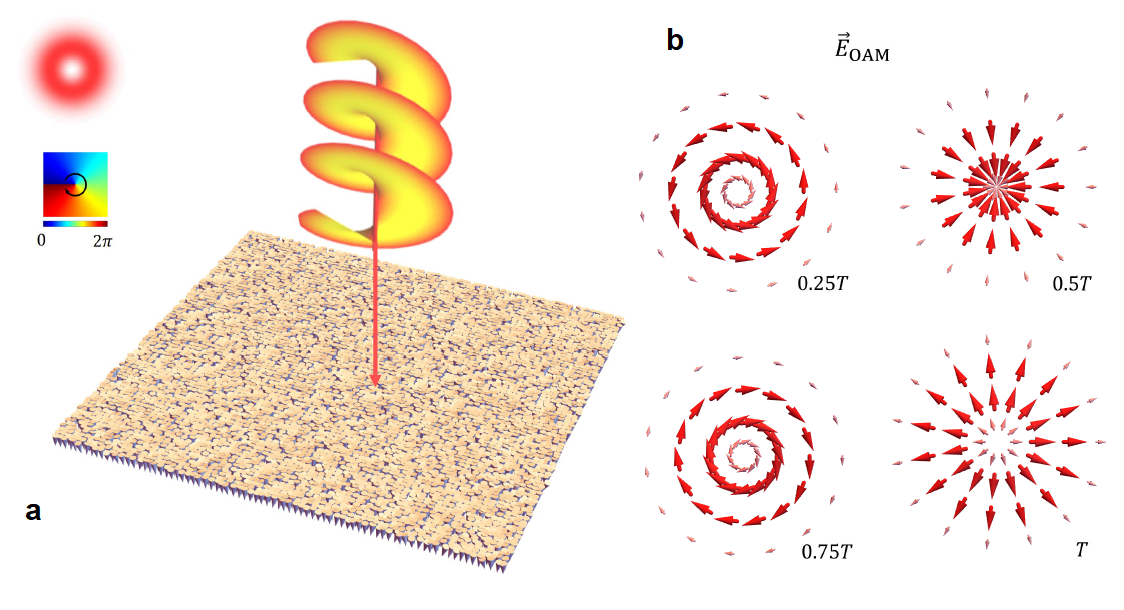}
\caption{\label{Figure1} \textbf{Illumination of poled PZT films by an ov beam}. \textbf{a} The helical wavefront represents the Ov beam. The red ``doughnut'' shape  in the upper left corner denotes the circularly-symmetric intensity profile of the $l=0$ and  $m=1$ ``Laguerre-Gaussian" mode, and the colorful palette shows its phase variation of $2\pi$. Local electric dipoles are represented by cones pointing down.  \textbf{b}  The evolution of the beam-generated electric fields within the film. The four panels show subsequent field configurations (anti-clockwise, convergent, clockwise and divergent vortices) separated by quarter of a period $T$.}
\end{figure}

Molecular dynamics simulations show that at the time $t = 3$ ps, the local dipoles already establish a well defined cyclical motion passing through a sequence of states shown in \textbf{Fig. 2a-b}. Each of such states has a continuous rotational symmetry around the central line of the vortex beam. Hence, in the subpanels of \textbf{Fig. 2a} we show the distribution of dipoles only in one of the radial cross sections, i.e. the $y=40$ $(x,z)$ plane passing through the rotational symmetry axis. Technically, the slab used to model our PZT films have eight (001) layers, including one for the substrate (layer 1), five for the film (layers 2-6), and two for the vacuum (layers 7 and 8). Additionally,  we schematically show in \textbf{Fig. 2a1, b1} the structural evolution of the dipoles in the top and bottom $(x,y)$ planes of the film (further plots showing the time evolution of dipole configurations at these two planes are reported in Fig. S2).

\begin{figure*} 
\includegraphics[width = 160mm]{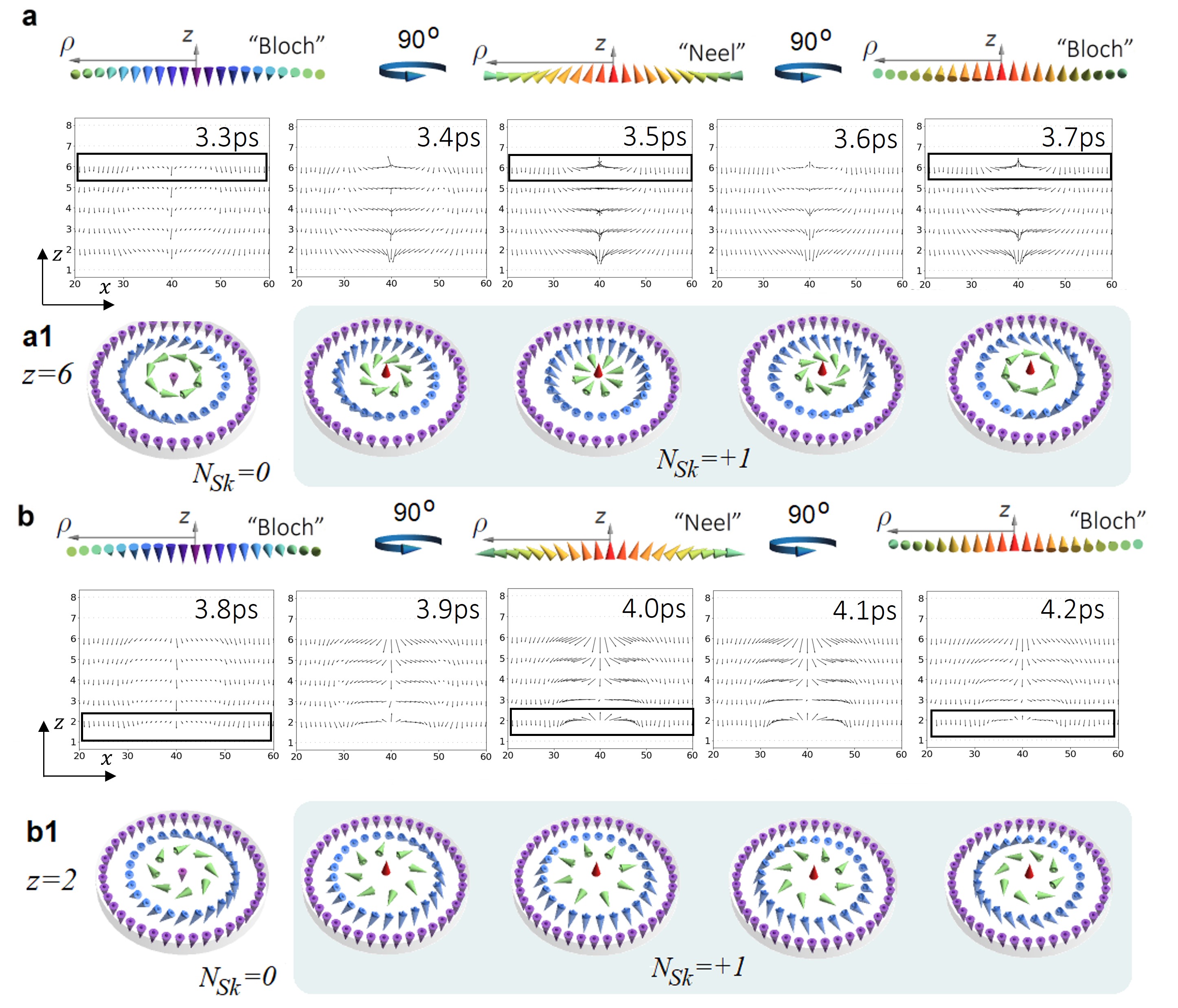} 
\caption{\label{Figure2} \textbf{Effective Hamiltonian simulation of the electric dipole evolution in the presence of an OAM field}. \textbf{a}. The view of the dipole configuration in the $y = 40$ plane in the first-half period ($3.3-3.7$ps), containing $P_x$ and $P_z$ (denoted by black arrows). The horizontal axis denotes the site number along the [100] direction and the vertical axis denotes the index of the (001) layers. Top row illustrates the ``Bloch"$\xrightarrow{}$``N\'{e}el"$\xrightarrow{}$``Bloch" rotations of the dipoles at the top layer ($z=6$) highlighted by the black box. \textbf{a1}. Schematic plot about the dynamical skyrmion evolution at the top interfacial layer during the first half period ($3.3-3.7$ps). \textbf{b}. Similar to \textbf{a}, the view of the dipole configuration in the $y = 40$ plane in the second-half period ($3.8-4.2$ps). Top row illustrates the ``Bloch"$\xrightarrow{}$``N\'{e}el"$\xrightarrow{}$``Bloch" rotations of the dipoles at the bottom layer ($z=2$) highlighted by the black box. \textbf{b1}. Schematic plot about the dynamical skyrmion evolution at the bottom interfacial layer during the second half period ($3.8-4.2$ps).  }
\end{figure*}

At $t = 3.3$ ps (\textbf{Fig. 2a}) the polar structure is homogeneous throughout all layers. The dipoles
 both at the center of the beam ($\rho=0$) and in the far-field region ($\rho>13$ u.c.) retain their original downward orientation. In contrast, at the intermediate distances from the central line ($1<\rho<10$ u.c.), the out-of-plane component ($P_z$) of electric dipoles is suppressed. Within this ring-shaped region, the in-plane components $P_x$ and $P_y$ grow following the intensity profile of the beam (\textbf{Fig. 1a}) and form an anti-clockwise vortex pattern (see the $0.25T$ field configuration in \textbf{Fig. 1b}). The rotations from the out-of-plane to in-plane orientations have a pronounced Bloch character as illustrated in the top leftmost panel of \textbf{Fig. 2a}. Overall, the structure at $t = 3.3$ ps can be described as an anti-clockwise annular vortex with downwards polarized core and surrounded by a downwards polarized matrix. 

As time passes, the downwards polarized core ($\rho=0$) rapidly undergoes a partial switching. For instance, at $t=3.4$ps the direction of the central dipole in top plane ($z=6$) is already reversed to $P_z>0$. At the same time, the core dipoles in all other planes are still oriented downwards but their magnitude decreases with increasing $z$. Such inhomogeneous core structure persists from $t\approx 3.38$ps up to $t\approx 3.75$ps, i.e. during almost a full half period. It is also important to note that owing to the gradient of $P_z$ at $\rho=0$, the magnitude of the central dipole in the $z=5$ plane almost vanishes at $t=3.5$ps and $t=3.6$ps. As will be discussed later, such  quasi-singular behavior as well as the reversal of $P_z$ at the top layer are non-trivial topological features. Note, however, that all layers possess similar patterns for the in-plane dipole components $P_x$ and $P_y$ at any given time (see Fig. S1).

Another structural change that occurs during the first half-period is the rotation of the dipoles around the $z-$axis (top row in \textbf{Fig. 2a}) which is remindful of the evolution of the OAM field (\textbf{Fig. 1b}). Such rotation gradually transforms the dominant Bloch component at $t=3.3$ps into its center-convergent N\'{e}el counterpart at $t=3.5-3.6$ps until, having accomplished a $180^\circ$ turn, the dipolar structure re-gains its assertive Bloch character at $t=3.7$ps, but is reversed with respect to $t=3.3$ps (rightmost vs leftmost top panel in \textbf{Fig. 2a}).

Thereby, the first half-period ends at $t=3.8$ps with an annular vortex state akin to the $t=3.3$ps structure but in a clock-wise manner. During the second half-period $t=3.8-4.2$ps (\textbf{Fig. 2b}), dipoles at each site continue to rotate anti-clockwise around the $z-$axis. Such rotations generate the center-divergent (e.g. $t=4.0$ps) and the anti-clockwise (e.g. $t=4.2$ps) vortex patterns in the ($x$,$y$) planes. The corresponding mutual transformation of Bloch and N\'{e}el rotations are schematically shown in the top row of \textbf{Fig. 2b}. Moreover, similar to the first-half-period evolution, the core polarization is partially switched during $t=3.88-4.25$ps. Yet, the reversal of $P_z$ is rather observed in the bottom ($z=2$) plane of the film, while in all other planes the magnitude of $P_z$ increases with $z$. Consequently, the quasi-singular point ($\textbf{P}\approx0$) occurs in the $z=3$ plane.

One can readily notice (\textbf{Figs. 2a1, b1}) that the optical vortex creates polar skyrmion textures at the top and bottom interfaces of the film whenever the direction of the central dipole in the corresponding layers is reversed. This observation is confirmed by the calculated evolution of the skyrmion number $N_{\rm{Sk}}$ for $z=2-6$ planes (\textbf{Fig. 3a}).
As the in-plane dipole components closely follow the morphology of the OAM field, the skyrmion helicity $\gamma$ also continuously evolves with time accompanying the vortex beam's phase, i.e. $\gamma=\omega t\,(\,\text{mod}\,2\pi)$~\cite{nagaosa2013topological}. For instance, a perfectly center-divergent (convergent) N\'{e}el skyrmion characterized by $\gamma=0$ ($\gamma=\pi$) forms at $t=T$ ($t=0.5T$) on the bottom (top) interface; in contrast, at the same time, center-divergent(convergent) skyrmions never occur at the top (bottom) plane (\textbf{Fig. 2a1, b1}). Additionally, at $t=0.25T$ or $t=0.75T$ ($\gamma=\pm\pi/2$), the system always opts for a topologically trivial annular vortex state instead of Bloch skyrmions.
Since the dipoles forming the skyrmion texture have to cover the full body angle, such behavior is topologically tied to the inhomogeneous switching of $P_z$ at the center line. 

Interestingly, such switching is not a direct effect of the beam-generated field since $\vec{E}(\vec{r},t)$ lacks an out-of-plane component at all times, but also vanishes at $\rho=0$. Instead, the switching mechanism roots in the electrostatic interactions between $P_{x,y}$ and $P_z$. Namely, the development of the radial component of polarization leads to a build-up of an electric bound charge $\rho_b^{x,y}=-(\partial_x P_x+\partial_y P_y)$. In response, by adopting a monotonically changing $P_z(z)$, the material develops a bound charge $\rho_b^z=-\partial_z P_z$  to compensate $\rho_b^{x,y}$ (illustrated in \textbf{Fig. 3b}). Such mechanism is further confirmed by our calculations of bound charges $\rho_b^z$ and $\rho_b^{x,y}$ evolving with time (see Supplementary Material). Thereby, depending on the sign and magnitude of $\rho_b^{x,y}$, the reversal of $P_z$ occurs either in the vicinity of the top or bottom interfaces whenever $d\gtrsim |{P_z}^{\rm{max}}|/|\rho_b^{z}|$, where $d$ denotes the film thickness. This condition also defines the critical values of  parameters (e.g. field magnitude $E_0$ and film thickness $d$) required to create polar skyrmion textures by the OAM light (see Supplementary Material).

The evolution of $P_z$ and $\rho_{\rm{b}}$ with time also allows to explain the formation of polar skyrmions from a topological perspective. Specifically, the continuity of polarization does not allow $N_{\rm{sk}}$ to vary with time, unless a topological transition occurs. This can be realized by introducing a Bloch point, a three-dimensional singularity with vanishing polarization. Such defect has been seen to mediate the dynamical evolution of magnetic skyrmions~\cite{zhou2015dynamically}. As being composed of planes with different $N_{\rm{sk}}$, Bloch point can induce a jump of $N_{\rm{sk}}$ by 1 when going through these planes. Above discussion about \textbf{Fig. 2} points out a negligible dipole moment at the center of layer 3 or layer 5 when skyrmion is present. Indeed, with the same method used in Ref.~\cite{thiaville2003micromagnetic}, by computing the topological charge of each cell, we find a negatively charged (-1) Bloch point at the center between layer 2 and layer 3 during 3.9--4.2 ps, and a positively charged (+1) Bloch point at the center between layer 5 and layer 6 during 3.4--3.7 ps, in concert with the presence of a skyrmion. According to \textbf{Fig. 2}, both Bloch points have spiraling configurations evolving with time as schematically shown in \textbf{Fig. 3c}, and they are involved in the creation of polar skyrmions.

We note $N_{\rm{sk}}$ of the two interfacial layers stay on ``0" and ``1" alternately. This behavior keeps proceeding as long as the OAM field does not vanish (Fig. S7). The interconversion between $N_{\rm{sk}}$ of ``0" or ``1" highly resembles digital binary signals and suggests that this dynamical skyrmion can be implemented as a component in logic gates. Though $N_{\rm{sk}}$ cannot be directly observed, it is associated with the switching of out-of-plane dipole components. This can be probed by measuring the out-of-plane polarization $P_z$ using interdigited electrodes~\cite{behera2022electric}. \textbf{Figure. 3d} shows how $P_z$ averaged within the beam focus at each (001) layer changes with time. With periodic evolution of dipole components, we see that when the corresponding N\'{e}el-type skyrmion is formed, $P_z$ at layer 2 or layer 6 reach its maxima and has positive values, in contrast to negative values at most other times. The sub-picosecond period represents a much faster switching process, as compared to conventional memory devices ($\mu$s) and magnetic skrymions (ns)~\cite{luo2021skyrmion}. Furthermore, the switching period of $N_{\rm{sk}}$ can be tuned by the frequency of the field (Fig. S8). As different from the motion of magnetic skyrmions driven by electric current~\cite{jiang2015blowing,woo2016observation}, here the nonlocal transport of polar skyrmion between two interfaces is realized via an electrostatic mechanism by applying an electric field, achieving the goal of less-energy dissipation~\cite{hsu2017electric}.

\begin{figure}
\includegraphics[width = 80mm]{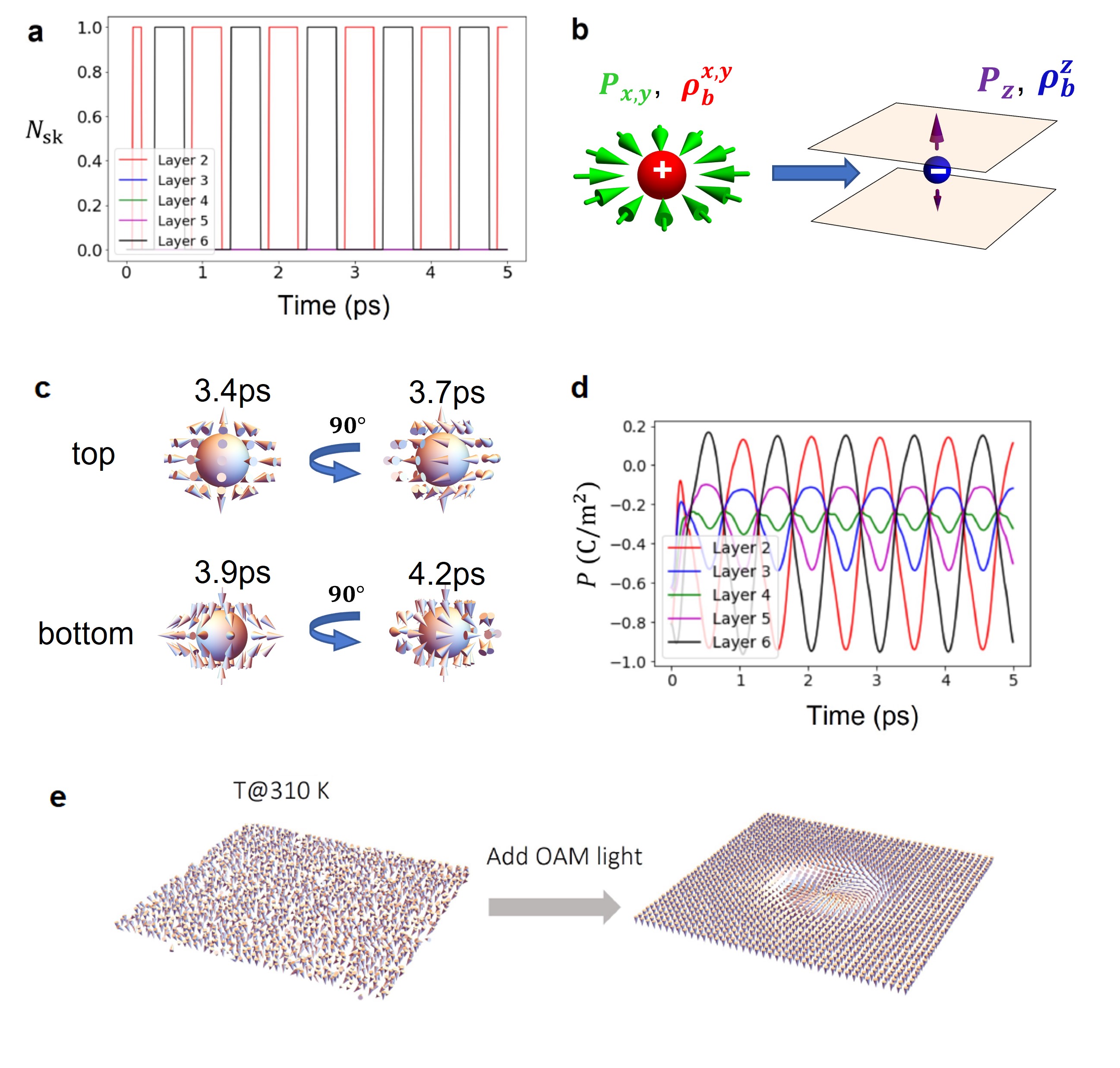}
\caption{\label{Figure4} \textbf{Topological and electrostatic origin of the OAM-induced-skyrmion}. \textbf{a} The variation of skyrmion number $N_{\rm{sk}}$ at each layer with time. \textbf{b} Illustration about bound charge density $\rho_b^{x,y}$ induced by $P_{x,y}$, and $\rho_b^{z}$ induced by $P_{z}$ to compensate $\rho_b^{x,y}$. \textbf{c} Illustration about two spiraling Bloch points located between layer 5 and layer 6 (first half period) and between layer 2 and layer 3 (second half period). 90 degree denotes dipole evolution with time. \textbf{d} The variation of averaged $P_z$ within beam focus at each layer with time. \textbf{e} OAM-induced-skyrmions survive even at room temperature. }
\end{figure}

Since electronic devices are usually operated at room temperature,  it is worthwhile to study how the OAM-induced-skyrmion behaves close to 300 K too. \textbf{Figure. 3e} shows the 3D plot of the monodomain configuration at 310 K before and after the illumination of an OAM field. Though dipoles fluctuate more at an elevated temperature, the skyrmion sustains within the focus spot out of the monodomain at 310K. As we vary the field magnitude and the screening of the depolarization field, the skyrmions are stabilized when  parameters span a large range (Fig. S5 and Fig. S9). All these evidences support the robustness of the mechanism and its practicality in application. By doubling the beam width, a larger skyrmion with a dimension$\sim$10 nm can be produced (Fig. S10), further demonstrating the high tunability of polar skyrmions governed by parameters of OAM field.

In  summary, we predict that the nontrivial winding pattern of the field in an optical vortex beam can be imprinted in ferroelectrics, and dipolar skyrmions will emerge and evolve dynamically out of an electrostatic cooperation between microscopic bound charges and external field. Effectively, engineering OAM optical field can be considered as the input `writing' process from one end, while the polarization patterns can correspond to the output `reading' process at the other end and being experimentally mapped by, e.g., a piezoelectric force microscopy (PFM) technique~\cite{han2022high}. This unusual light-matter interaction delivers a new perspective about designing fast-speed memory and logic devices.

We thank Max Mignolet, Dr. Peng Chen, Dr. Liuyang Sun, Hao Song, and Prof. Alan E. Willner for useful discussions. We acknowledge the support from the Grant MURI ETHOS W911NF-21-2-0162 from Army Research Office (ARO) and the Vannevar Bush Faculty Fellowship (VBFF) Grant No. N00014-20-1-2834 from
the Department of Defense.  We also acknowledge the computational support from the Arkansas High Performance Computing Center for computational resources. 

\bibliography{reference.bib}

\foreach \x in {1,...,22}
{
\clearpage
\includepdf[pages={\x,{}}]{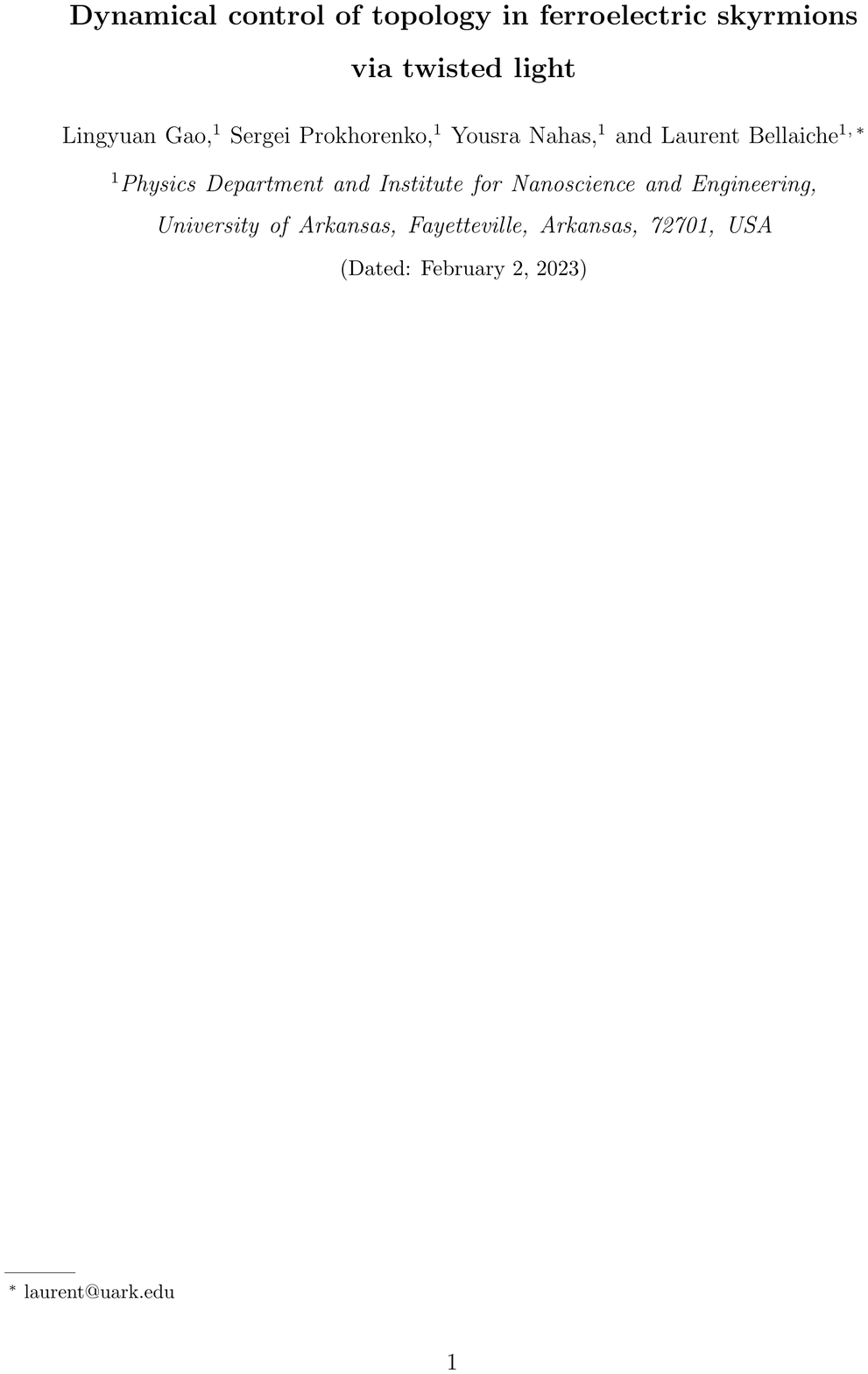}
}

\end{document}


\preprint{}

\title{Dynamical control of topology in ferroelectric skyrmions \\ via twisted light}

\author{Lingyuan Gao}
\author{Sergei Prokhorenko}
\author{Yousra Nahas}
\author{Laurent Bellaiche}
\email{laurent@uark.edu}
\affiliation{Physics Department and Institute for Nanoscience and Engineering, University of Arkansas, Fayetteville, Arkansas, 72701, USA}

\date{\today}

\maketitle

The supplementary material contains six sections:

(i) Computational methods;

(ii) A brief review about Laguerre-Guassian (LG) modes and orbital angular momentum (OAM) of light;

(iii) Computational details about effective Hamiltonian simulations;

(iv) Electrostatic origin of OAM-induced ferroelectric skyrmions;

(v) Critical conditions and robustness of OAM-induced-skyrmions;

(vi) Characterizing topology of defects. 

\section{Computational methods}
We model Pb(Zr$_{0.4}$Ti$_{0.6}$)O$_{3}$ (PZT) ferroelectric ultrathin films grown along the [001] direction with a Pb-O layer as the terminated interfaces/surfaces. An $80 \times 80 \times n$ supercell with periodicity along the [100] and [010] pseudo-cubic directions but being finite along the pseudo-cubic [001] direction is used. Under a compressive strain $\sim$-2$\%$ accounting for the lattice mismatch between substrate and PZT films, and an 88$\%$ screening of the polarization-induced surface charge resembling the electrodes effect, a monodomain configuration with electric dipoles aligning along [001] direction can be established. Technically, the first-principles-based effective Hamiltonian approach is used in Monte Carlo or Molecular Dynamics simulations  to determine energetics and local electric dipoles of each 5-atom perovskite unit cell (u.c.)~\cite{zhong1994phase,zhong1995first}, and here we adopt the letter. The validity of this approach has been demonstrated in previous studies about PZT ultrathin films, including (1) predicting the formation of 180$^{\circ}$ up and down stripe domains along in-plane direction as the ground state~\cite{kornev2004ultrathin, lai2006electric}, and discovering a linear dependency between the width of stripes and the square root of film thickness~\cite{lai2007thickness}; these theoretical predictions agree well with experimental observations~\cite{streiffer2002observation, schilling2006scaling}; (2) predicting the emergence of the dipolar maze (labyrinthine) and nano-bubble (skyrmionic) phases as non-equilibrum, self-assembled states, which is also supported by experimental evidences~\cite{nahas2020topology}; (3) predicting an inverse transition from the maze pattern to the less-symmetric parallel-stripe pattern with the elevated temperature, as consistent with experimental findings too~\cite{nahas2020inverse}. In addition, various types of topological defects in ferroelectrics such as  bubbles~\cite{lai2006electric}, vortices~\cite{naumov2004unusual}, dipolar waves~\cite{sichuga2011epitaxial} and merons~\cite{lu2018topological} have been predicted by this method and confirmed experimentally~\cite{sichuga2011epitaxial,zhang2017nanoscale,lu2018topological}. To run molecular dynamics simulations, we gradually cool down the system from 650 K to 10 K at an interval of 20 K. At each temperature, we run for 50000 steps with a time step of 0.5 fs to reach thermal equilibrization. At 10 K, a monodomain configuration forms for the ground state, and all electric dipoles are aligned out-of-plane and point downwards. We then introduce the optical vortex beam at the well-equilibrated monodomain at this temperature.

\section{Laguerre-Gaussian modes and orbital angular momentum of light}
Here, we briefly review the derivation of LG modes, and the concept of OAM of light. According to Maxwell's equations, the electromagnetic (EM) wave is usually described by three-dimensional wave equations:
\begin{equation}
    \nabla^{2}\psi(\Vec{r}) + k^2 \psi(\Vec{r}) = 0,
    \label{s1}
\end{equation}
where $\nabla^{2}$ is the Laplacian operator, $k$ is the wavenumber,  and $\psi({\Vec{r}})$ can be related to the electric field or magnetic field by multiplying the unit vector $\vec{e}$ of the polarized field  as $\Vec{E}(\Vec{r}) = \Vec{e} \psi(\Vec{r})$ and $\Vec{B}(\Vec{r}) = \Vec{e} \psi(\Vec{r})$. In cylindrical coordinates ($\rho, \phi, z$),  $\rho$, $\phi$ and $z$ denote the radial coordinate, azimuthal angle and the coordinate along the cylindrical axis, respectively. Note that, for a wave propagating along the cylindrical axis,  $\vec{k} \parallel \vec{z}$, and $\psi(\Vec{r}) = u(\Vec{r})e^{ikz}$. With that, Eq.~\ref{s1} is rewritten as:
\begin{equation}
    (\nabla^2_{T} + 2ik\frac{\partial}{\partial z} + \frac{\partial^2}{\partial z^2})u(\Vec{r}) = 0,
    \label{s2}
\end{equation}
where $\nabla^2_{T} = \frac{\partial^2}{\partial \rho^2} + \frac{1}{\rho}\frac{\partial}{\partial \rho} + \frac{1}{\rho^2}\frac{\partial^2}{\partial \phi^2}$ is the transverse component of the Laplacian operator. Using paraxial approximation:
\begin{equation}
    |\frac{\partial^2 u}{\partial z^2}| \ll |\frac{\partial^2 u}{\partial x^2}|, \,
    |\frac{\partial^2 u}{\partial z^2}| \ll |\frac{\partial^2 u}{\partial y^2}|, \,
    |\frac{\partial^2 u}{\partial z^2}| \ll 2k|\frac{\partial u}{\partial z}|,
    \label{s3}
\end{equation}
the third term within the bracket of Eq.\ref{s2} can be dropped, and the most general form of LG mode is given by~\cite{andrews2011structured}:
\begin{align}
    u(\rho, \phi, z) = \frac{C_l^{|m|}}{w(z)}\Big(\frac{\sqrt{2}\rho}{w(z)} \Big)^{|m|}
    e^{-\frac{\rho^2}{w^{2}(z)}} L_l^{|m|}\Big{(}\frac{2\rho^2}{w^{2}(z)}\Big{)}e^{im\phi}e^{i\varphi(z)}e^{-ik\frac{\rho^2}{2R(z)}},
    \label{s4}
\end{align}
where $w(z) = w_0 \sqrt{1 + (\frac{z}{z_R})^2}$ is the beam width, $w_0$ is the beam waist and $z_R$ is the Rayleigh length; $R(z)=z(1 + (z_R/z)^2)$ represents the radius of curvature of the beam's wavefront at $z$; $\varphi(z)$ is called the Guoy phase; $C_l^{|m|}$ is the normalized coefficient; $L_l^{|m|}$ is the generalized Laguerre function. The mode profile $|u(\rho, \phi, z)|^2$ is circularly symmetric and highly depends on two integer parameters $l$ and $m$, but is independent of the polarization vector $\Vec{e}$. Throughout the main text, we take $l = 0$ and $m = 1$. Also, since the film is very thin, we take $z = 0$ and $\varphi(z) = 0$. Considering all these conditions, without writing the constant coefficient explicitly, the $l = 0$ mode profile can be written as:
\begin{equation}
    u(\rho, \phi, z) = u_0 \Big(\frac{\sqrt{2}\rho}{w} \Big)^{|m|}
    e^{-\frac{\rho^2}{w^{2}}} e^{im\phi}.
    \label{s5}
\end{equation}
Associating Eq.\ref{s5} with the polarization vector of a circularly-polarized light $\Vec{e} = \Vec{e_x} + i\Vec{e_y}$, we get Eq. 1 in the main manuscript as the form of the electric field.

Let us also indicate that, as analogous to the definition of angular momentum $\Vec{L} = \Vec{r} \times \Vec{p}$ of classical objects, the total angular momentum of electromagnetic (EM) wave $\vec{L}^{\rm{tot}}$ can be defined as:
\begin{equation}
    \Vec{L}^{\rm{tot}} = \epsilon_0 \int \Vec{r} \times (\Vec{E} \times \Vec{B}) d^{3}\Vec{r} =  \epsilon_0 \int \Vec{r} \times \big( \Vec{E} \times (\nabla \times \Vec{A}) \big) d^{3}\Vec{r},
    \label{s6}
\end{equation}
where the Coulomb gauge $\nabla \cdot \Vec{A} = 0$ is adopted. $\vec{L}^{\rm{tot}}$ can be written as the sum of two parts: one part is related to the unit vector of polarization $\vec{e}$, called ``spin angular momentum" (SAM) $\vec{L}^{\rm{S}}$; for a left- or right-polarized light, $\vec{L}^{\rm{S}} = \pm \hbar$. The other part depends on the mode profile $u(\rho, \phi, z)$, and is called ``orbital angular momentum" (OAM) $\vec{L}^{\rm{O}}$. 

Under paraxial approximation, for a monochromatic wave, $\vec{L}^{\rm{O}}$ is given by~\cite{belinfante1940current}:
\begin{equation}
    \Vec{L}^{\rm{O}} = \epsilon_0 \sum_{i = x, y, z}\int E_i (\Vec{r} \times \nabla)A_{i} d^{3}\Vec{r} = \frac{\epsilon_0}{2i\omega} \sum_{i = x, y, z}\int {E_i}^{*} (\Vec{r} \times \nabla)E_{i} d^{3}\Vec{r}.
    \label{s7}
\end{equation}
In cylindrical coordinates, the $z$-component of $\vec{L}^{\rm{O}}$ is:
\begin{equation}
    L^{\rm{O}}_{z} = \frac{\epsilon_0}{2i\omega} \sum_{i = x, y, z}\int {E_i}^{*} \frac{\partial E_{i}}{\partial \phi} d^{3}\Vec{r},
    \label{s8}
\end{equation}
which is the OAM of light. For $l = 0$ mode with a profile $u(\rho, \phi, z) \propto e^{im\phi}$,
\begin{equation}
    L^{\rm{O}}_{z} = m\hbar.
    \label{s9}
\end{equation}

\section{Computational details for effective Hamiltonian model simulations}

As introduced in the main manuscript, we set $\omega = 1$ THz in our calculations corresponding to a period $T$ = 1 ps for the field, since dynamics of electric dipoles are typically  at a time scale of picoseconds.  We choose $w$ = 5 u.c. considering the dimension of the system and confine the field within a circular boundary in any $(x, y)$ plane. Such optical field is promising to be realized in experiments, considering the facts that (1) the focus spot of THz field can shrink down to nano-size with a largely enhanced field intensity~\cite{rusina2008nanoconcentration,schnell2011nanofocusing,toma2015squeezing}, thanks to the development of nanoantenna arrays and plasmonic waveguides\cite{gramotnev2010plasmonics,novotny2011antennas}, (2) OAM of light can also be transferred to plasmonic vortex beams and generated at nanoscale size~\cite{heeres2014subwavelength,pu2015near,garoli2016optical,spektor2017revealing,hachtel2019spatially}, and (3) the beam’s wave front and phase can be manipulated by locally engineering nanoantenna arrays~\cite{genevet2012ultra,williams2013optical,karimi2014generating,jin2021phyllotaxis}. Here, we set $E_0 = 3 \times 10^{10}$ V/m, noting that the magnitude of computed electric fields is often overestimated by a factor of about 20 in effective Hamiltonian approaches with respect to measurements~\cite{xu2017designing, jiang2018giant}; also in accordance with the exponential factor in Eq. (1) of the main text, the field drops rapidly along the outward radial direction. Experimentally, ferroelectric films with ultrathin thickness ($\sim$nm) is feasible to withstand an electric field of the order of $10^9$ V/m~\cite{garcia2009giant,chanthbouala2012solid,garcia2014ferroelectric}. Moreover, the dielectric constant of PZT significantly drops at THZ frequency range~\cite{kwak2011dielectric}, naturally implying that a much higher breakdown strength should occur for a high-frequency optical field~\cite{mcpherson2002proposed} than for DC fields ($\sim 10^8$ V/m)~\cite{ko2019improvement} , further confirming that our simulated fields are realistic. Note also that the magnetic part of the EM field is not considered in the present simulations, since (i) in SI units, for an electromagnetic wave, the magnetic field is 10$^{8}$ times smaller than the electric field; and (ii) PZT is a nonmagnetic material with a spin splitting that is small even in the presence of magnetic field.
 
\section{Electrostatic origin of the dynamical polar skyrmion evolution}
Here, we scrutinize the mechanism behind  (1) the fact that, while for any given time the pattern of OAM field is identical between the five different (001) layers, these latter adopt different morphologies for the $z$-component of the dipolar pattern, while  in-plane dipole components have similar qualitative patterns between layers (see Fig. S1)  and (2) the striking change of these dipolar morphologies with time (see Fig. S2). For that, the left column of Fig.~\ref{FigureS3}(a) shows the morphology of the OAM field in any $(x,y)$ plane at four different times. The field does not have any out-of-plane component, and $E_x$ and $E_y$ are  identical  for all five layers at each of these four times -- with this morphology changing between times. The intensity of the field at the center of any (001) layer, that is at $\rho = 0$, is zero. It increases with $\rho$ and reaches a maximum at $\rho_{\rm{max}} = \frac{w}{\sqrt{2}}$, at any time.  At $t$ = 3.5 ps and $t$ = 4.0 ps, the convergent or divergent form of the electric field acts as an external charged source or sink sitting at the center of layers, respectively. In fact, as we are going to show, bound charge is induced there.

To demonstrate that, we calculated the bound charge density as $\rho_{\rm{b}} = -\nabla \cdot \bm{P}$ at the same times as those in Fig.~\ref{FigureS3}(a) and plot it in Fig.~\ref{FigureS3}(b). 
It is important to know that our numerical results show that the variance between $\rho_{\rm{b}}$ at different (001) layers is small at any chosen time (see Fig.~\ref{FigureS33}), so we thus only plot $\rho_{\rm{b}}$ at layer 3 in Fig.~\ref{FigureS3}(b). It is also worthwhile to  realize that, since the non-uniformity of both in-plane and out-of-plane dipole components contribute to the induced bound charge, the total bound charge density $\rho_{\rm{b}}^{\rm{total}}$ can be divided into $\rho_{\rm{b}}^{xy}$ induced by both $P_x$ and $P_y$ {\it versus} $\rho_{\rm{b}}^{z}$ that is solely originating from $P_z$.

In the right column of Fig.~\ref{FigureS3}(a), we note that the pattern of in-plane dipole components follows the morphology of the OAM field, though with a time delay less than 0.05 ps. This is owing to the intrinsic energy coupling $-\bm{p} \cdot \bm{E}$ which dictates that the lowest-energy configuration of dipoles is the parallel alignment between dipoles and fields~\cite{bellaiche2001electric}, with the delay being due to a finite response time of dipoles to the external field -- especially if the involved configurations are inhomogeneous in nature ~\cite{ponomareva2008nature}, as in the present case. As shown in Fig.~\ref{FigureS3}(b), at $t$ = 3.25 ps and 3.75 ps, the solenoidal form of both OAM field and in-plane dipole components only induce a small $\rho^{xy}_{\rm{b}}$. Consequently, the induced $\rho^{z}_{\rm{b}}$ is also small, and that corresponds to a more uniform pattern of the $z$-component of the electric dipoles along the $z$ direction close to the center of each layer at $t$ = 3.3 ps and 3.8 ps, as shown in Fig. 2 in the main manuscript. In contrast, at $t$ = 3.5 ps and 4.0 ps,  a large positive and negative $\rho^{xy}_{\rm{b}}$ emerges at the center of layers from the divergent or convergent in-plane dipole configuration, respectively. As a response, $P_z$ adjusts its arrangement so that the induced $\rho^z_{\rm{b}}$ is opposite to $\rho^{xy}_{\rm{b}}$, in order to lower the total bound charge density $\rho^{\rm{total}}_{b}$. In other words, the presence of the external electric field induces the redistribution of macroscopic bound charges associated with the $x$- and $y$-components of the electric dipoles, which in turn acts upon the $z$-component of the electric dipoles to counteract such redistribution. This provides a mechanism to the formation of polar skyrmions at the bottom ($z=2$) and top ($z=6$) layers.

\section{Critical conditions and robustness of OAM-induced skyrmions}

Witnessing the critical role that $\rho_{\rm{b}}$ plays in the formation of skyrmions, we wonder whether we can directly manage $\rho_{\rm{b}}$ so as to control skyrmion generation. Since $\rho_{\rm{b}}$ depends on the variation of in-plane dipole components $P_x$ and $P_y$, which are strongly coupled to the external field $E_{\rm{OAM}}$, we examine OAM fields at different magnitudes (see Fig.~\ref{FigureS4}), with respect to the above results of $E_0 = 30\times 10^9$ V/m. A field with relatively small magnitude, e.g., $E_0 = 6 \times 10^{9}$ V/m, drives a pattern with small in-plane components of the electric dipoles within the focus spot at each (001) layer, and only a small $\rho^{xy}_{\rm{b}}$ is induced without any inversion of $P_z$ in any (001) layer. When $E_0$ increases to $12\times 10^{9}$ V/m, the augmented in-plane components of the dipole induce a large $\rho^{xy}_{\rm b}$, which in turn leads to flipped $P_z$ at layer 2. If we keep increasing $E_0$ to the range of $50-100 \times 10^{9}$V/m, $\rho_{\rm{b}}$ does not change much from $\rho_{\rm{b}}$ at $E_0 = 30\times 10^9$ V/m, and skyrmions are robust when $E_0$ varies in such a large range.

The monotonous change of $P_z$ at the center of each layer along the [001] direction displayed in Fig. 2 poses another interesting question whether there is a critical thickness for the dipole inversion. This is because the $P_z$ at the center of one interfacial layer can be reversed with respect to $P_z$ at the center of other (001) layers only after ``walking through'' enough distance along the [001] direction. Indeed, we find that, for $N = 3$, the magnitude of $P_z$ at the center of layers decreases from the top to the bottom layer of the film, but its sign at the center of the bottom layer is not switched with respect to that of the top layer (see Fig.~\ref{FigureS5}).
In contrast, by adding one more layer (that is considering $N = 4$), we see the recovery of the dipole inversion at the bottom (001) layer at 4.0 ps, indicating that $N_{\rm{cr}} = 4$ is the critical thickness (see Fig. S5). Consistent with the value of $N_{\rm{cr}}$, the dipole inversion holds in thicker films $N$ = 6, or 7 (see Fig.~\ref{FigureS5}), and this feature informs that polar skyrmions can be observed in thin films with thickness larger or equal than $N_{\rm{cr}} = 4$.

Figure.~\ref{FigureS6}-~\ref{FigureS9} show the robustness of skyrmions generated out of this mechanism under various conditions by varying intrinsic parameters of the OAM field, including the pulse duration, optical frequency, screenings to polarization and beam radius.

\section{Characterizing topology of defects}
To characterize the topology of ferroelectric skyrmion, we calculate the skyrmion number $N_{\rm{sk}}$ of each (001) layer by integrating the Pontryagin charge density $\rho_{\rm{sk}}$ over the full plane:
\begin{equation}
    N_{\rm{sk}} = \int \rho_{\rm{sk}}(\vec{r}) d^2 \vec{r} =  \int \frac{1}{4\pi}\vec{n}(\vec{r})\cdot\big( \frac{\partial \vec{n}(\vec{r})}{\partial x} \times \frac{\partial \vec{n}(\vec{r})}{\partial y} \big) d^2 \vec{r},
    \label{s10}
\end{equation}
where $\vec{n}(\vec{r})$ is the normalized local mode vector at site $\vec{r}$.

To compute the monopole charge of a Bloch point in each cell, we follow the numerical approach introduced in Ref.~\cite{motrunich2004emergent} , which is mathematically equivalent to approaches used in Ref.~\cite{lau1989numerical,thiaville2003micromagnetic}. We choose two normalized local mode vectors $\vec{n}_i$, $\vec{n}_j$ on two neighbouring sites out of eight corners of a cube and another reference vector $\vec{n}_{*}$ to form a spherical triangle ($\vec{n}_{*}$, $\vec{n}_{i}$, $\vec{n}_{j}$). With the solid angle $\Omega[\vec{n}_{*}, \vec{n}_{i}, \vec{n}_{j}]$ subtended by this spherical triangle, an auxiliary variable gauge potential $A_{ij}$ can be defined as:
\begin{align}
    e^{i A_{ij}} = e^{\frac{i}{2}\Omega[\vec{n}_{*}, \vec{n}_{i}, \vec{n}_{j}]}=\frac{1 + \vec{n}_{*}\cdot\vec{n}_{i} + \vec{n}_{*}\cdot\vec{n}_{i} + \vec{n}_{*}\cdot\vec{n}_{j} + i \vec{n}_{*}\cdot( \vec{n}_{i} \times \vec{n}_{j})}{\sqrt{2(1 + \vec{n}_{*}\cdot\vec{n}_{i})(1 + \vec{n}_{*}\cdot\vec{n}_{j})( 1 + \vec{n}_{i}\cdot\vec{n}_{j})}}.
\end{align}
Using $A_{ij}$, the flux $F$ on the face bounded by site (1,2,...n,1) can be defined as:
\begin{equation}
    e^{iF} =  e^{i(A_{12} + A_{23} + ... + A_{n1})},
\end{equation}
and $F \in (-\pi, \pi]$. Note $F$ is gauge invariant and is independent of the choice of the reference vector $\vec{n}_{*}$. In practice, we compute $F$ for triangle faces defined by three corners on the cube. By summing $F$ over all twelve triangles covering the volume of cube (illustrated in Fig.~\ref{FigureS11}), the monopole charge is computed as:
\begin{equation}
    k = \frac{\sum_{i=1}^{12} F_i}{2\pi},
\end{equation}
which is an integer. 

\clearpage

\begin{figure*}
\includegraphics[width = 160mm]{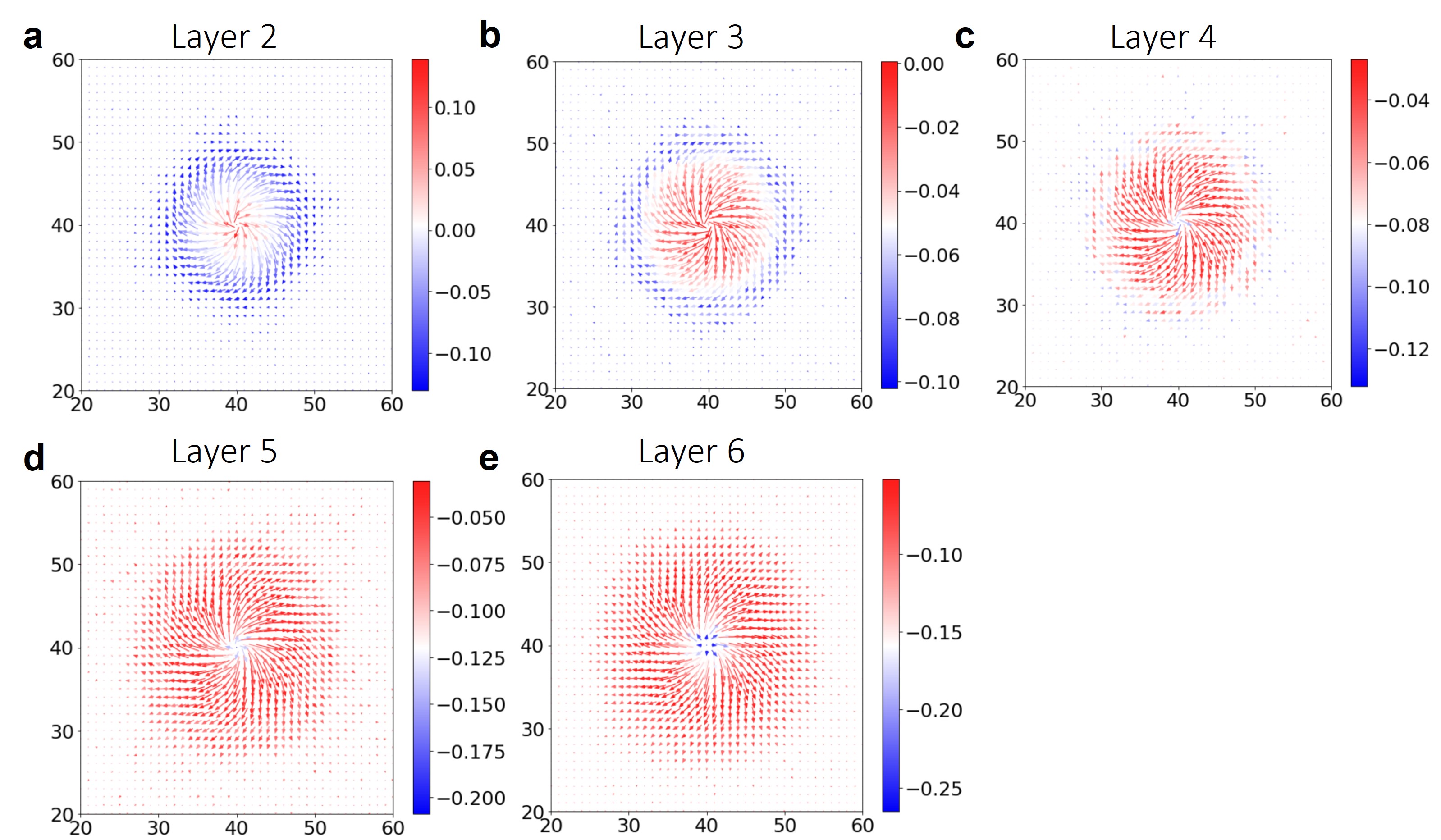}
\caption{\label{FigureS1} \textbf{2D plot for dipole components on layers 2-6 at t = 4.0ps}. The $P_x$ and $P_y$ are denoted by arrows and $P_z$ (arbitrary unit) is denoted by the colorbar. Horizontal and vertical axes denote site numbers along the [100] and [010] directions, respectively. }
\end{figure*}

\clearpage

\begin{figure}
\includegraphics[width = 160mm]{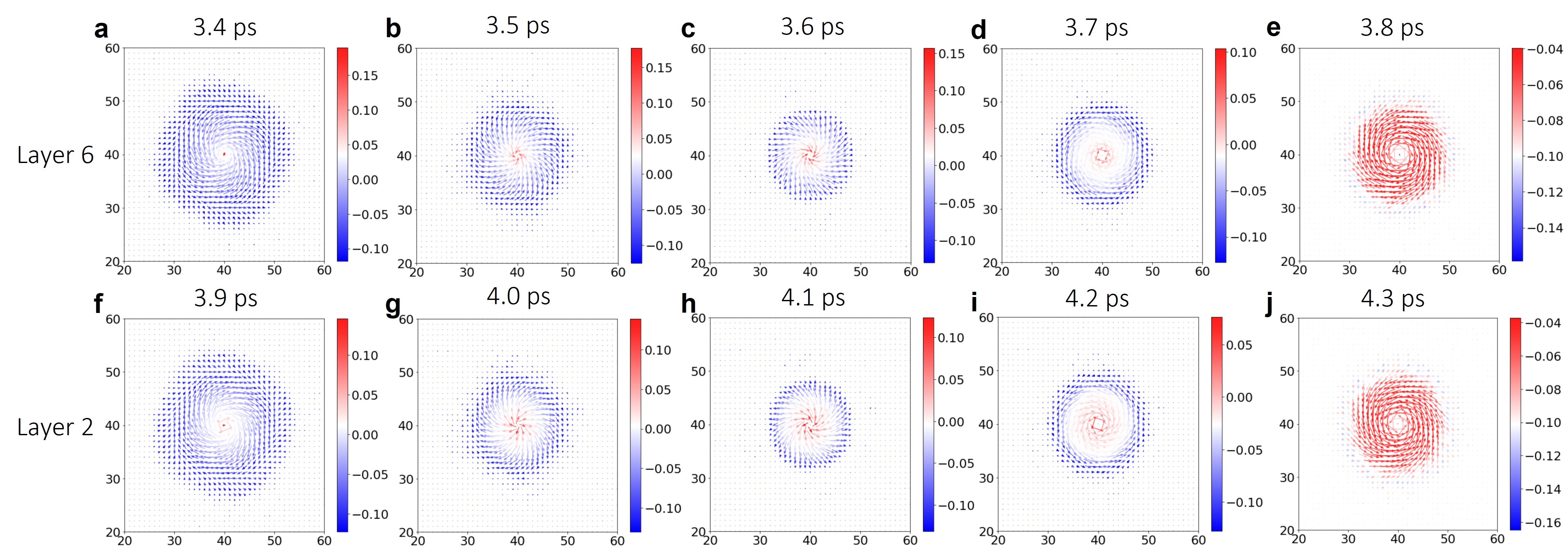}
\caption{\label{FigureS2} \textbf{2D plot for dipole components on layer 2 and layer 6 at different times (3.4-4.3 ps).} \textbf{a-e} are for layer 6 and \textbf{f-j} are for layer 2.}
\end{figure}

\clearpage

\begin{figure}
\includegraphics[width = 160mm]{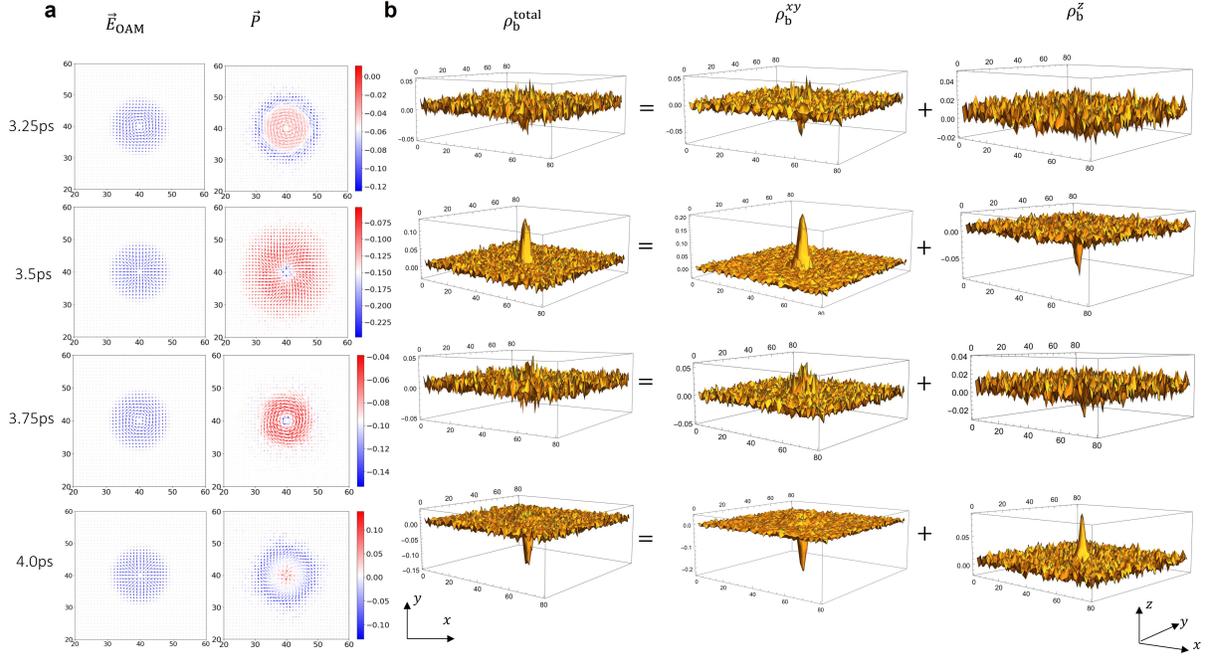}
\caption{\label{FigureS3}
\textbf{The electrostatic origin of the skyrmion generation induced by OAM field}. \textbf{a} The morphology of the OAM field (first column) and the dipole pattern on layer 2 (second column) at time 3.25 ps, 3.5 ps, 3.75 ps and 4.0 ps, respectively. Both fields and dipoles are in arbitrary units. The horizontal and vertical axes denote the site number along the [100] and [010] directions respectively, and the OAM field as well as the $x$- and $y$-components of the electric dipoles  are shown by arrows. The $z$-component of these dipoles is represented by the colorbar in arbitrary unit. \textbf{b} The bound charge density $\rho_{\rm{b}}$ (in arbitrary unit) induced by the OAM field at time 3.25 ps, 3.5 ps, 3.75 ps and 4.0 ps in layer 3. The first column is the total $\rho_{\rm{b}}$, the second column is $\rho_{\rm{b}}$ induced by in-plane dipole components $P_x$ and $P_y$ and the third column is $\rho_{\rm{b}}$ induced by out-of-plane dipole component $P_z$. The in-plane axes denote the site number along [100] and [010] directions, and the vertical axis denotes $\rho_{\rm{b}}$ in arbitrary unit.
}
\end{figure}

\begin{figure}
\includegraphics[width = 160mm]{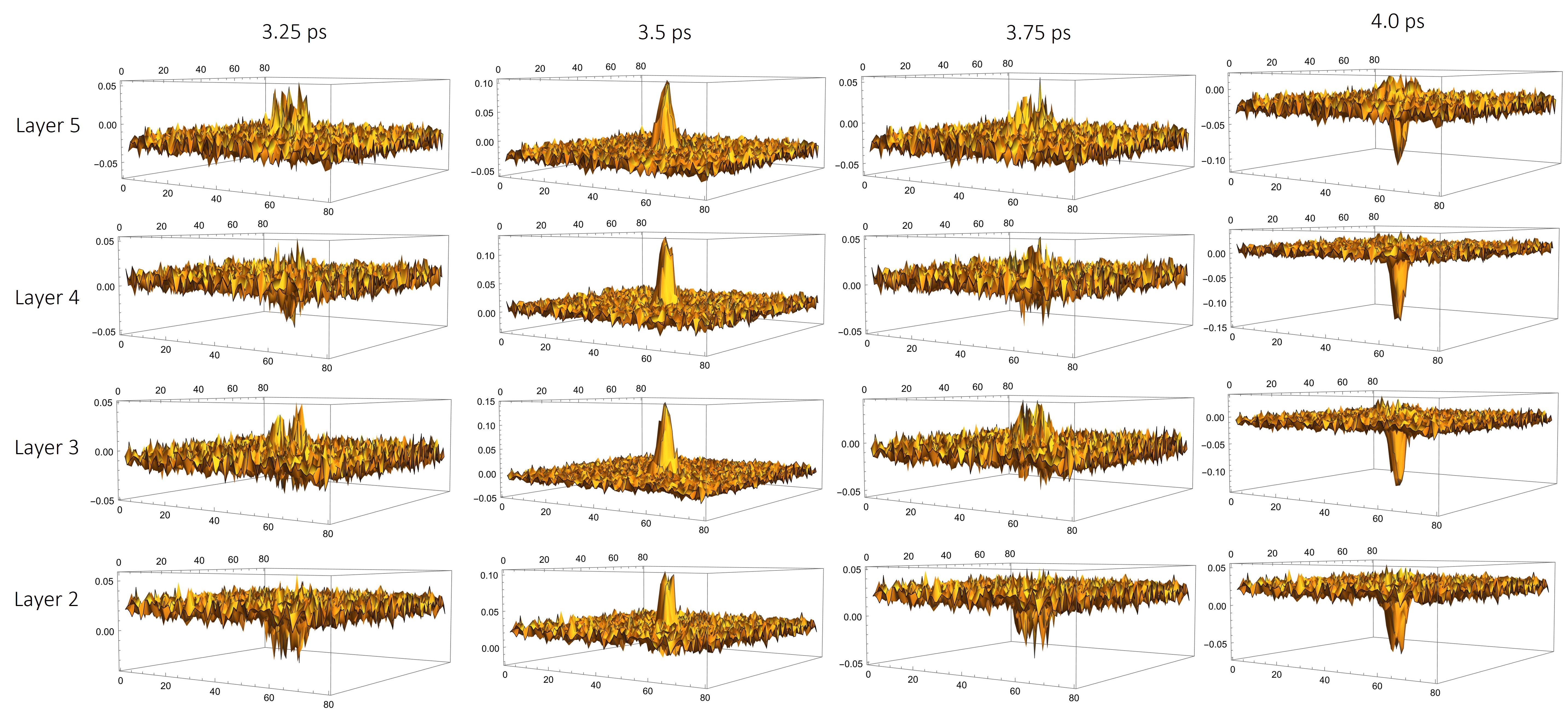}
\caption{\label{FigureS33} \textbf{The total bound charge density $\rho^{\rm{total}}_b$ on layers 2-5 at four different times: 3.25 ps, 3.5 ps, 3.75 ps, and 4.0 ps.}  Horizontal and vertical axes denote site numbers along the [100] and [010] directions, respectively, and $z$- axis denotes $\rho^{\rm{total}}_b$ in arbitrary unit. }
\end{figure}

\clearpage

\begin{figure}
\includegraphics[width = 160mm]{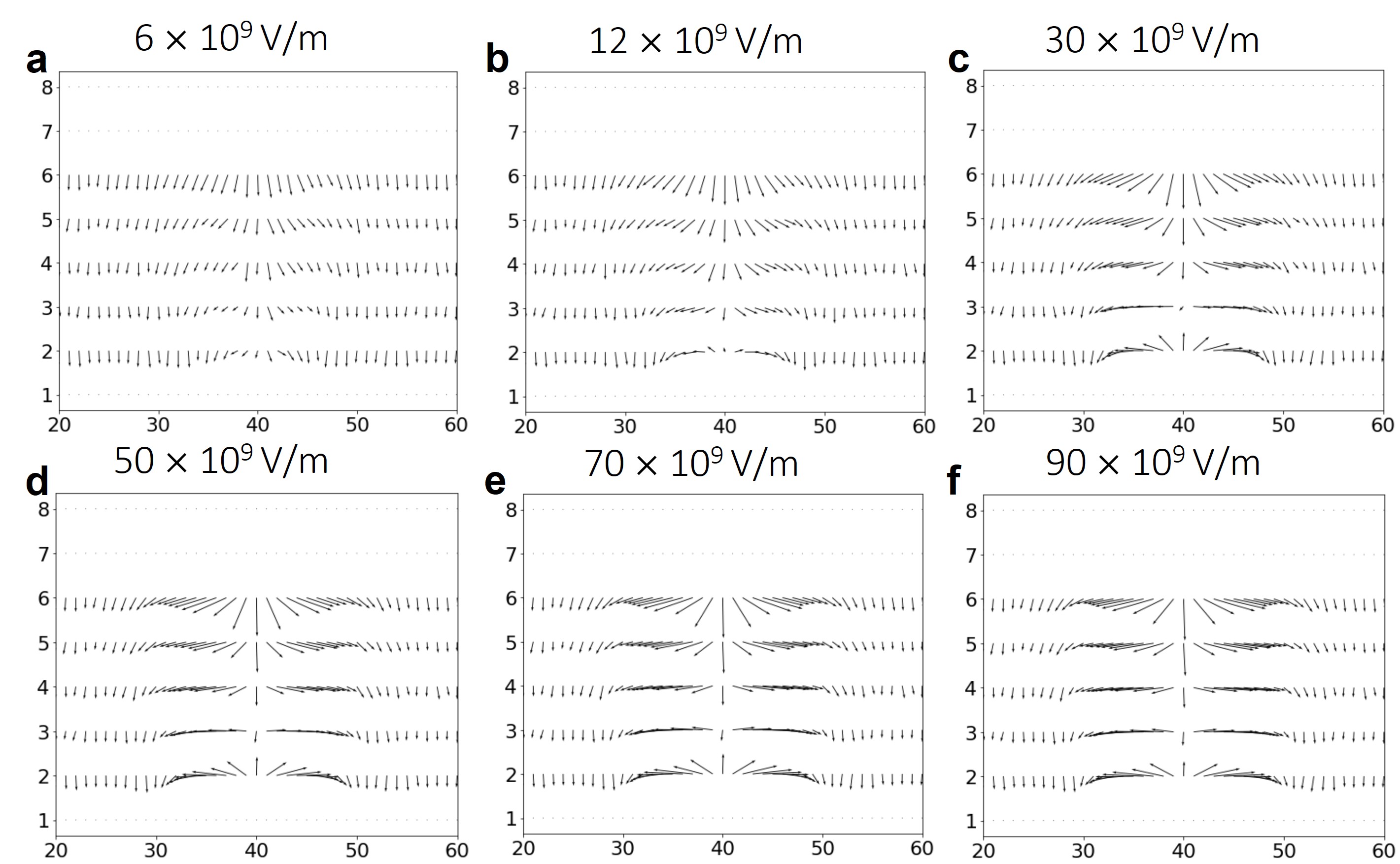}
\caption{\label{FigureS4} \textbf{Side view of the dipole configuration in the $y = 40$ plane at t = 4.0 ps, under OAM fields with different magnitudes}. $P_x$ and $P_z$ are contained in this plane and are denoted by black arrows. The horizontal axis denotes the site number along [100] direction and the vertical axis denotes the index of the (001) layers. \textbf{a}-\textbf{f} correspond to $E_0$ = 6, 12, 30, 50, 70, 90 $\times 10^{9}$ V/m, respectively. }
\end{figure}

\clearpage

\begin{figure}
\includegraphics[width = 160mm]{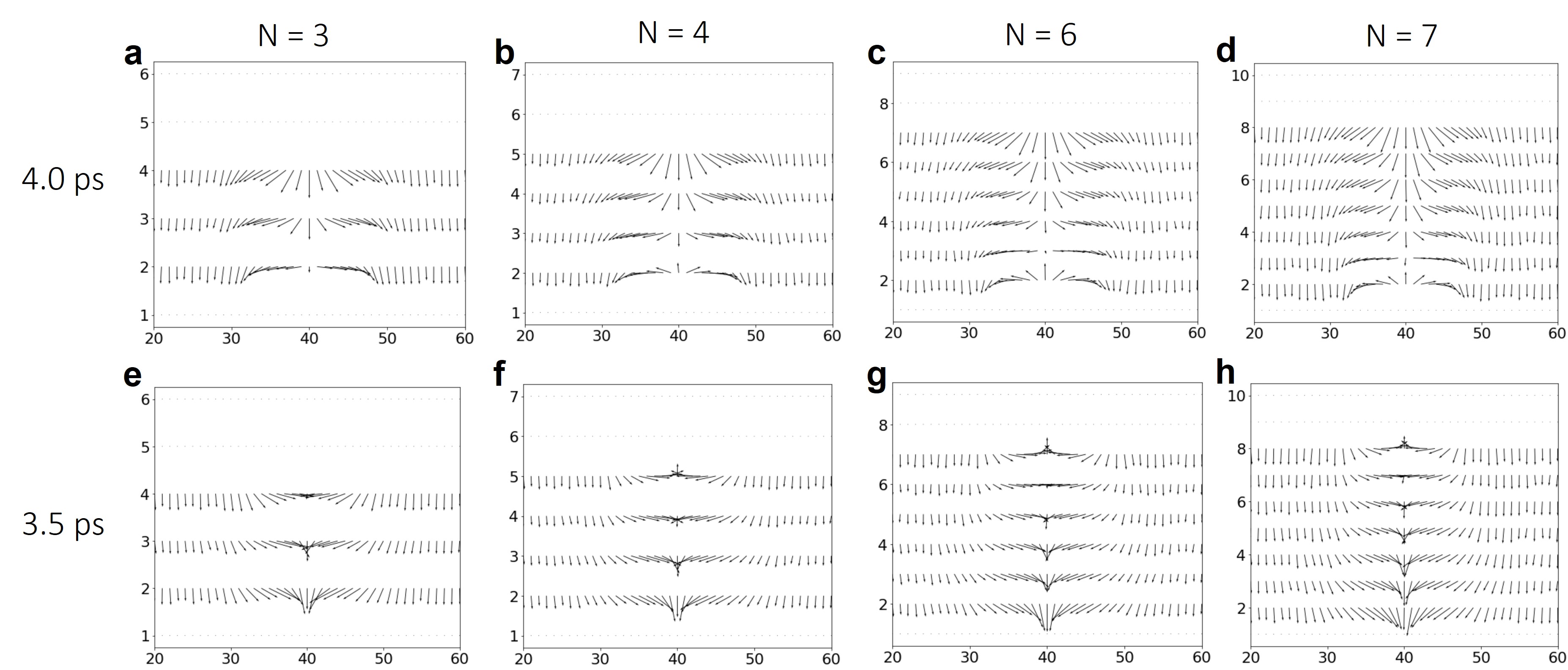}
\caption{\label{FigureS5} \textbf{Side view of the dipole configuration in the $y = 40$ plane, for systems with different thickness}. Similar to Fig. S4, $P_x$ and $P_z$ are contained  in this plane. \textbf{a} and \textbf{e}, \textbf{b} and \textbf{f}, \textbf{c} and \textbf{g}, and \textbf{d} and \textbf{h} correspond to a film thickness N = 3, 4, 6, 7, respectively. The first line (\textbf{a}-\textbf{d}) and the second line (\textbf{g}-\textbf{h}) represent snapshots at t = 4.0 ps and t = 3.5 ps, respectively.   }
\end{figure}

\clearpage

\begin{figure}
\includegraphics[width = 160mm]{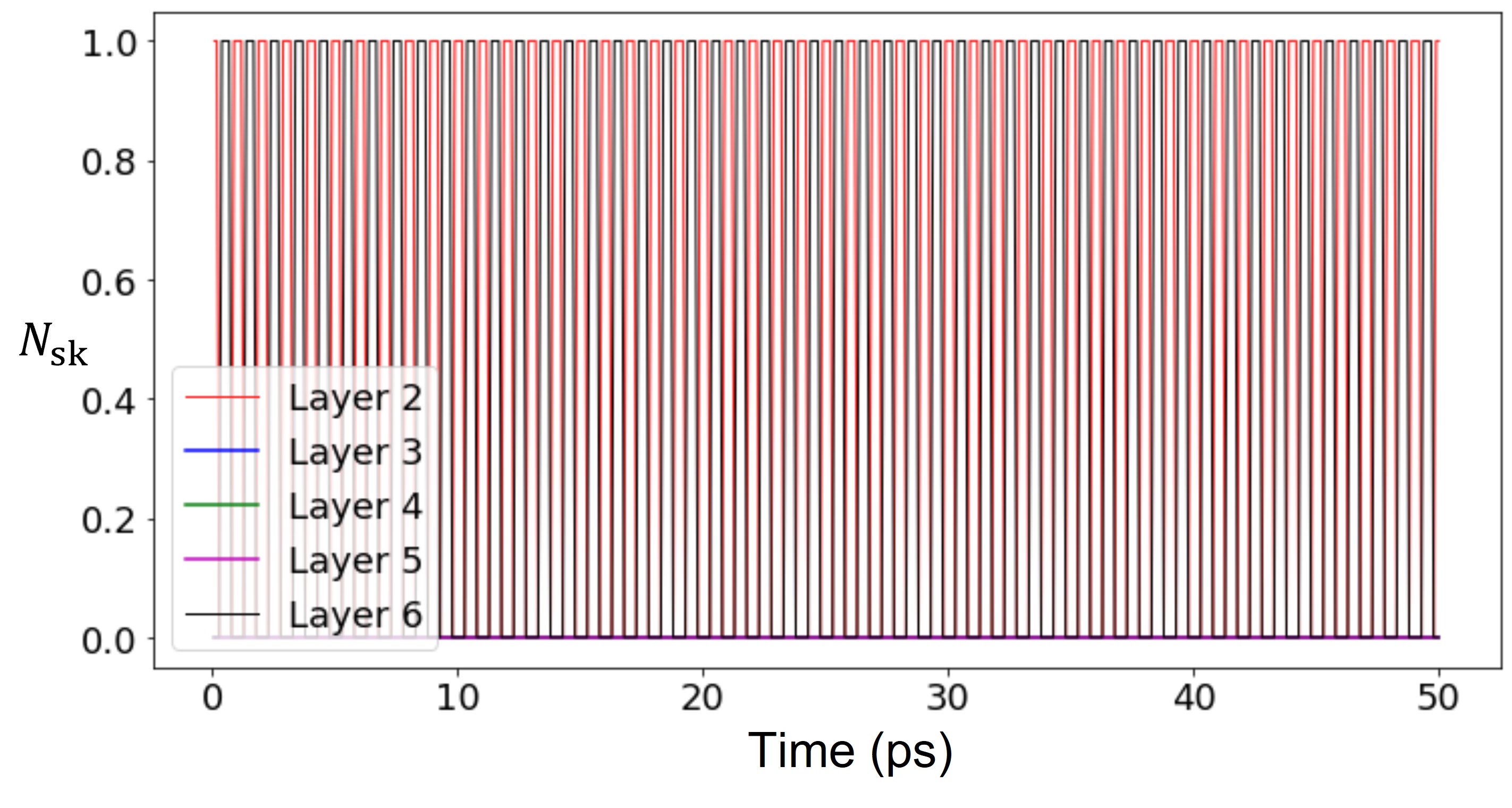}
\caption{\label{FigureS6} \textbf{A long calculation (50 ps) showing the alternating skyrmion number $N_{\rm{sk}}$ of the system with time in the presence of the OAM field. }}
\end{figure}

\clearpage

\begin{figure}
\includegraphics[width = 160mm]{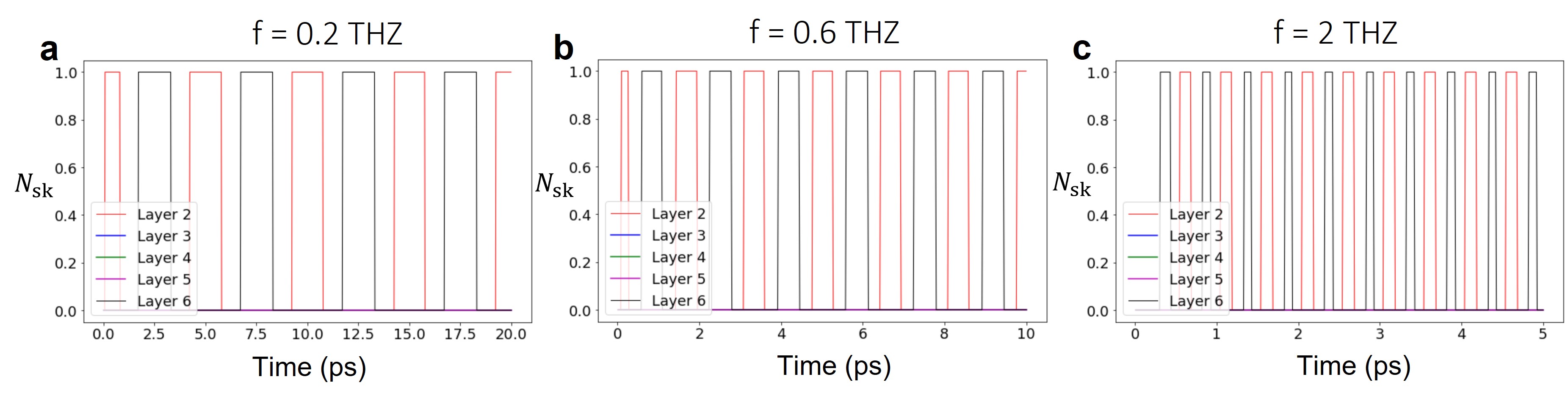}
\caption{\label{FigureS7} \textbf{Different alternating periods of $N_{\rm{sk}}$ tuned by frequencies of the OAM field}. \textbf{a}-\textbf{c} represent f = 0.2 THZ, 0.6 THZ and 2 THZ, respectively.}
\end{figure}

\clearpage

\begin{figure}
\includegraphics[width = 160mm]{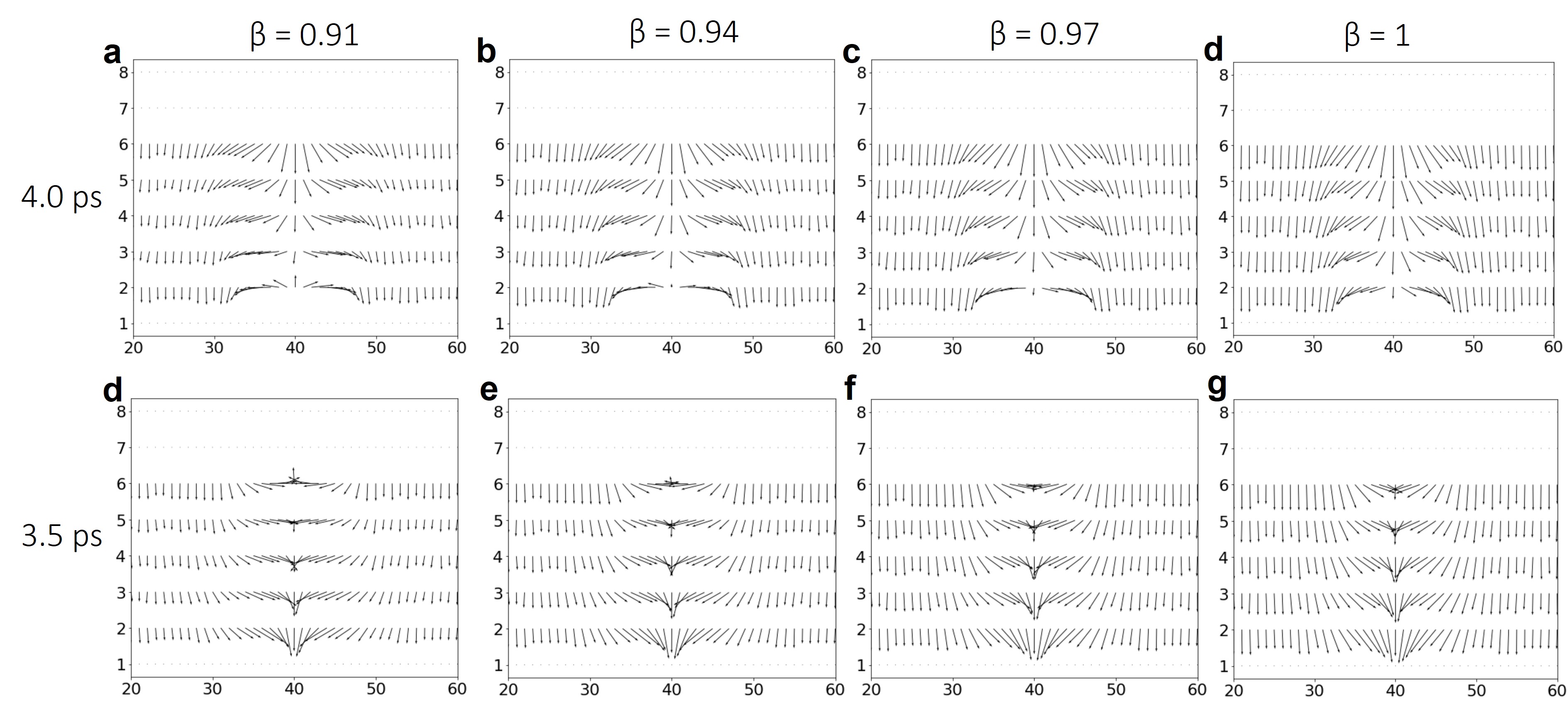}
\caption{\label{FigureS8} \textbf{Side view of the dipole configuration in the $y = 40$ plane, for systems with different $\beta$ parameters denoting the screening to the polarization-induced surface charge.} \textbf{a} and \textbf{d}, \textbf{b} and \textbf{e}, \textbf{c} and \textbf{f} and \textbf{d} and \textbf{g} represent $\beta$ = 0.91, 0.94, 0.97 and 1, respectively. The first line (\textbf{a}-\textbf{d}) and the second line (\textbf{g}-\textbf{h}) represent snapshots at t = 4.0 ps and t = 3.5 ps, respectively.  }
\end{figure}

\clearpage

\begin{figure}
\includegraphics[width = 160mm]{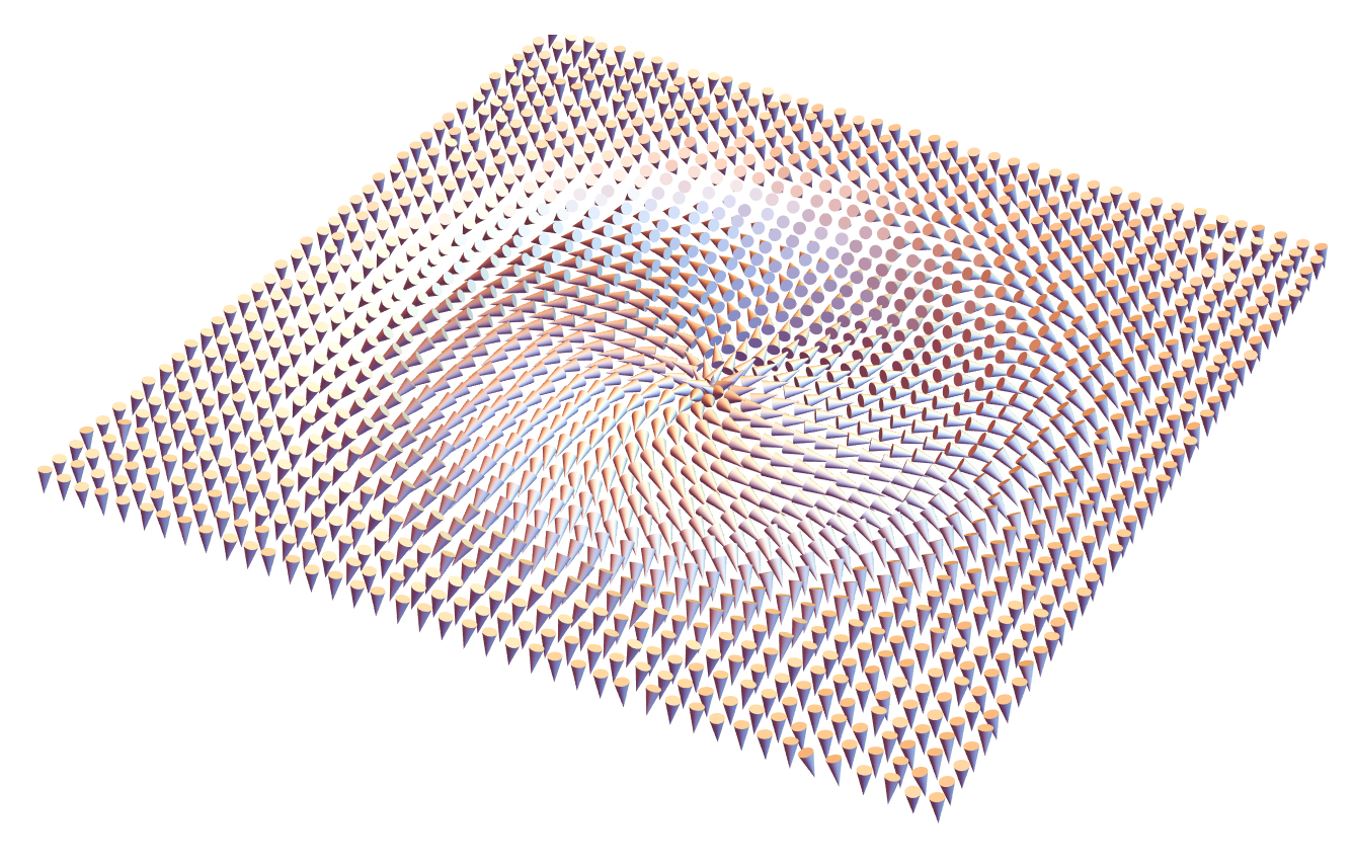}
\caption{\label{FigureS9} \textbf{A larger skyrmion of the diameter $\sim$10 nm, induced by OAM field with the beam radius of $w = 10$ u.c. and $E_0 = 18 \times 10^{10}$ V/m}.}
\end{figure}

\clearpage

\begin{figure}
\includegraphics[width = 40mm]{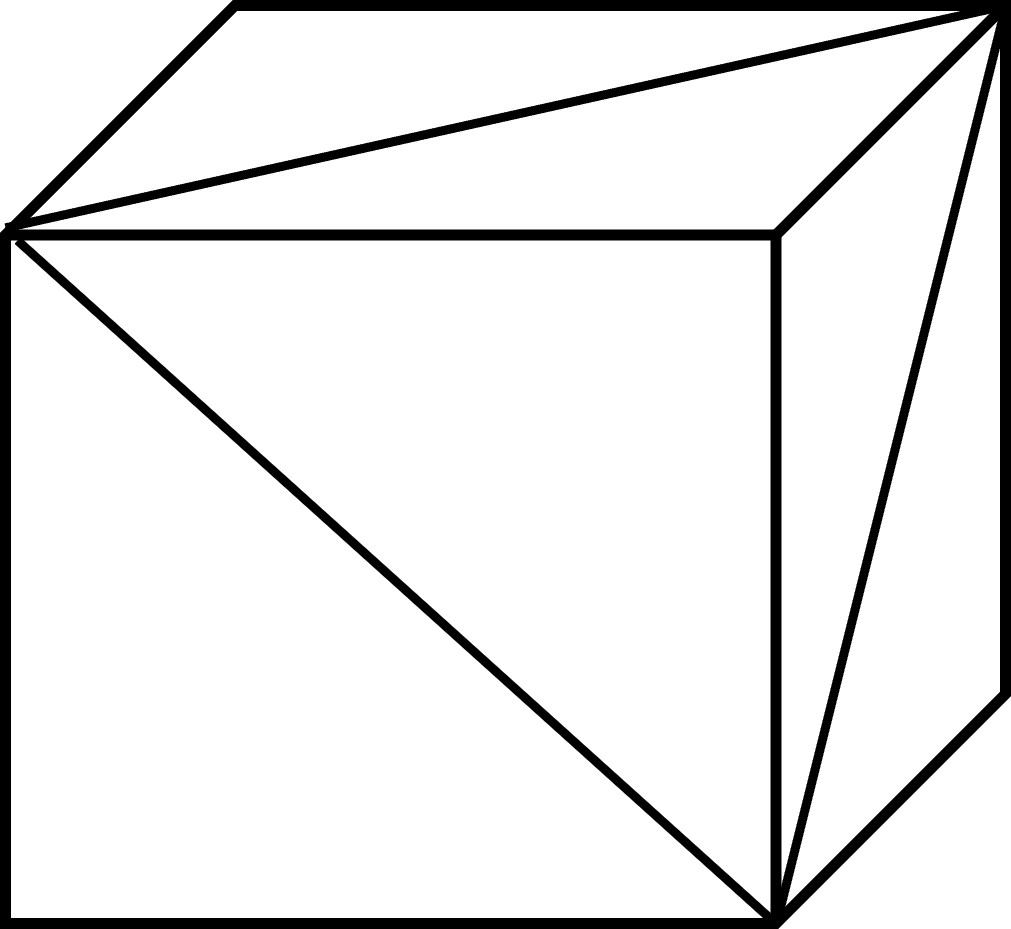}
\caption{\label{FigureS11} \textbf{Illustration about computing monopole charge of a Bloch point in each cell}. We compute the flux of each triangle face denoted with three corners ($i,j,k$) on a cube by using local mode at these sites. We then sum over the flux of twelve triangles covering the cubic volume to compute the monopole charge.}
\end{figure}

\clearpage

\bibliography{reference.bib}